\documentclass[twocolumn]{aastex62}

\usepackage{natbib}

\begin{document}

\title{AGB interlopers in YSO catalogues hunted out by NEOWISE}

\author[0000-0003-3119-2087]{Jeong-Eun Lee}
\affil{School of Space Research, Kyung Hee University,
1732, Deogyeong-daero, Giheung-gu, Yongin-si, Gyeonggi-do 17104, Korea\\ :\href{mailto:jeongeun.lee@khu.ac.kr}{jeongeun.lee@khu.ac.kr}}

\author{Sieun Lee}
\affil{School of Space Research, Kyung Hee University,
1732, Deogyeong-daero, Giheung-gu, Yongin-si, Gyeonggi-do 17104, Korea\\ :\href{mailto:jeongeun.lee@khu.ac.kr}{jeongeun.lee@khu.ac.kr}}

\author{Seonjae Lee}
\affil{Department of Physics and Astronomy, Seoul National University, 1 Gwanak-ro, Gwanak-gu, Seoul 08826, Korea}

\author[0000-0001-9104-9763]{Kyung-Won Suh}
\affil{Department of Astronomy and Space Science, Chungbuk National University, Cheongju-City, 28644, Korea}

\author{Se-Hyung Cho}
\affil{Department of Physics and Astronomy, Seoul National University, 1 Gwanak-ro, Gwanak-gu, Seoul 08826, Korea}
\affil{Korea Astronomy and Space Science Institute (KASI), 776 Daedeokdae-ro, Yuseong-gu, Daejeon 34055, Korea}

\author{Do-Young Byun}
\affil{Korea Astronomy and Space Science Institute (KASI), 776 Daedeokdae-ro, Yuseong-gu, Daejeon 34055, Korea}
\affil{University of Science and Technology, 217 Gajeong-ro, Yuseong-gu, Daejeon 34113, Korea}

\author{Wooseok Park}
\affil{School of Space Research, Kyung Hee University,
1732, Deogyeong-daero, Giheung-gu, Yongin-si, Gyeonggi-do 17104, Korea\\ :\href{mailto:jeongeun.lee@khu.ac.kr}{jeongeun.lee@khu.ac.kr}}

\author{Gregory Herczeg}
\affil{Kavli Institute for Astronomy and Astrophysics, Peking University, Yiheyuan 5, Haidian Qu, 100871 Beijing, China}
\affil{Department of Astronomy, Peking University, Yiheyuan 5, Haidian Qu, 100871 Beijing, China}

\author{Carlos Contreras Pe{\~n}a}
\affil{Centre for Astrophysics Research, University of Hertfordshire, Hatfield AL10 9AB, UK}
\affil{School of Physics, Astrophysics Group, University of Exeter, Stocker Road, Exeter EX4 4QL, UK}

\author[0000-0002-6773-459X]{Doug Johnstone}
\affil{NRC Herzberg Astronomy and Astrophysics, 5071 West Saanich Road, Victoria, BC, V9E 2E7, Canada}
\affil{Department of Physics and Astronomy, University of Victoria, 3800 Finnerty Road, Elliot Building, Victoria, BC, V8P 5C2, Canada}

\begin{abstract}
AGBs and YSOs often share the same domains in IR color-magnitude or color-color diagrams leading to potential mis-classification. We extracted a list of AGB interlopers from the published YSO catalogues using the periodogram analysis on NEOWISE time series data. 
YSO IR variability is typically stochastic and linked to episodic mass accretion. Furthermore, most variable YSOs are at an early evolutionary stage, with significant surrounding envelope and/or disk material.
In contrast, AGBs are often identified by a well defined sinusoidal variability with periods of a few hundreds days.
From our periodogram analysis of all known low mass YSOs in the Gould Belt, we find 85 AGB candidates, out of which 62 were previously classified as late-stage Class III YSOs.  Most of these new AGB candidates have similar IR colors to O-rich AGBs. We observed 73 of these AGB candidates in the H$_2$O, CH$_3$OH and SiO maser lines to further reveal their nature. The SiO maser emission was detected in 10 sources,  confirming them as AGBs since low mass YSOs, especially Class III YSOs, do not show such maser emission. The H$_2$O and CH$_3$OH maser lines were detected in none of our targets.

\vspace{10mm}
\end{abstract}


\vspace{10mm}

\section{Introduction} \label{sec:intro} 

Both forming and dying stars are bright at infrared (IR) wavelengths since their effective temperatures are as low as 2000 to 4000 K, and they are often enshrouded by cold and dense circumstellar material. As a result, forming stars, ``young stellar objects (YSOs)", as well as dying stars, ``evolved stars", are commonly identified via their location in IR color-magnitude or color-color diagrams \citep{Suh11, Tu13, Koenig14}. However, the colors of YSOs and evolved stars, especially asymptotic giant branch stars (AGBs), overlap significantly in these diagrams producing potential contamination of source identifications. 

Gaia astrometry can distinguish between YSOs and AGB stars towards regions of diffuse molecular clouds and modest extinctions \citep{Manara18,Herczeg19}, however this separation is usually not possible for sources in very dense clouds with active star-formation.
Precise classification of AGBs can be performed based on the mid-IR (MIR) spectral features formed by their circumstellar dust \citep{Sylvester99, Suh14}. Unfortunately, it is difficult to carry out such MIR spectroscopic observations from the ground.
Alternatively, especially for the pulsating AGBs (with typical periods of 100--1000 days), we can use their light curves as a discriminator because the dynamics of pulsation is very regular and produces a distinctly periodic light curve \citep{Hofner18}. 

YSOs are also variable due to time-dependent mass accretion  from their circumstellar disks to their central forming stars. However, most YSO accretion variability on timescales shorter than a few years is rather stochastic because of the nature of the accretion process (Park et al. submitted). In some cases, however, even protostars show quasi-periodic variability probably because of interactions with companions or long-lived structures in the inner disk. For example, an embedded protostar EC 53 (as known as V371 Ser and Ser SMM5) varies in IR \citep{Hodapp96, Horrobin97} and submillimeter \citep{Yoo17} with a period of $\sim$1.5 years, 
possibly because a forming planet very close to the central protostar induces a burst accretion every 1.5 years \citep{Hodapp12, Lee20}. 
Only a few other protostars are known to have regular, pulsed accretion events \citep[]{Muzerolle13}, in some cases related to known binarity \citep{Tofflemire17}, and usually on week-months timescales. Hot or cold spots in the rotating young stellar photosphere also generate regular sinusoidal light curves, but with much shorter periods of days \citep[e.g.][]{Cody14, Sergison20}.

Most AGB stars are long-period variables (LPVs) with large amplitude pulsations. LPVs are classified into small-amplitude red giants (SARGs), semi-regular variables (SRVs), and Miras according to the characteristics (amplitude, period, and regularity) of the pulsation \citep{Soszynski13}. An AGB star is believed to evolve from a SARG to an SRV and finally to
a Mira variable, increasing its pulsating period and amplitude. 
SRVs have periods between $\sim$125 and $\sim$175 days and amplitudes smaller than 2.5 mag in the V-band, while Miras show regular periods peaked around 275 days and amplitude greater than 2.5 mag in the V-band \citep{Kholopov96}.

In this paper, using the MIR light curves of NEOWISE, we identify AGB interlopers from established YSO catalogues \citep{Dunham15, Megeath12, Esplin19} for nearby low mass star forming regions (Section 2) and describe H$_2$O, CH$_3$OH, and SiO maser observations toward those AGB candidates to confirm their nature (Section 3). We report on and discuss the results of our maser observations in Section 4 and Section 5, respectively.

\section{\textit{WISE}/NEOWISE variable samples} \label{sec:wise}

The Near-Earth Object Wide Infrared Survey Explorer (NEOWISE) provides all sky survey photometric data at 3.4 (WISE band 1; W1) and 4.6 (WISE band 2; W2) $\mu$m every six months \citep{Cutri15}. We adopted the NEOWISE data released on March 2020, which includes the data from the first 6 years (12 epochs) between 2014 and 2019. Previously, these time series MIR photometric data have been used to compare variability at both MIR and submillimeter wavelengths for a concise sample of 59 YSOs \citep{Contreras20}. Park et al. (submitted) analysed further the NEOWISE data set to identify variable YSOs found in established  YSO catalogs of nearby low mass star forming regions  based on Spitzer and Herschel data \citep{Dunham15, Megeath12, Esplin19}.\footnote{In this paper we use the source numbers from the original YSO catalogs to identify sources, appending ``M" and ``D" for the \citet{Megeath12} and \citet{Dunham15} catalogs, respectively.} Although a significant fraction of embedded protostars (Class 0 and I YSOs) and disk sources (Class II YSOs) are found to vary at IR wavelengths, almost all of this observed variability is either stochastic or  monotonically ascending and descending brightness changes with timescales longer than 6 years. 

In contrast, about one third of the catalogued late-stage variable YSOs (proposed as Class III) have sinusoidal light curves with periods of a few hundreds days and high fractional amplitudes (Park et al. submitted). Only 1--2\% of variable YSOs in earlier evolutionary stages 
show such regular sinusoidal light curves.
This regular variability with periods between 200 and 1000 days and high fractional amplitude is unexpected for YSOs, especially for the Class III YSOs. Figure 1 presents the MIR light curves of the W2 magnitude for two NEOWISE variable sources with identified periods shorter than 1000 days.\footnote{See Park et al.~submitted for the detailed procedure to produce these light curves from the NEOWISE photometric data and to determine the periodic variability with the periodogram analysis.} 

 \begin{figure}[t]
\centering
{\includegraphics[trim={0.5cm 0.3cm 0.3cm 0.3cm},clip,width=0.98\columnwidth]
{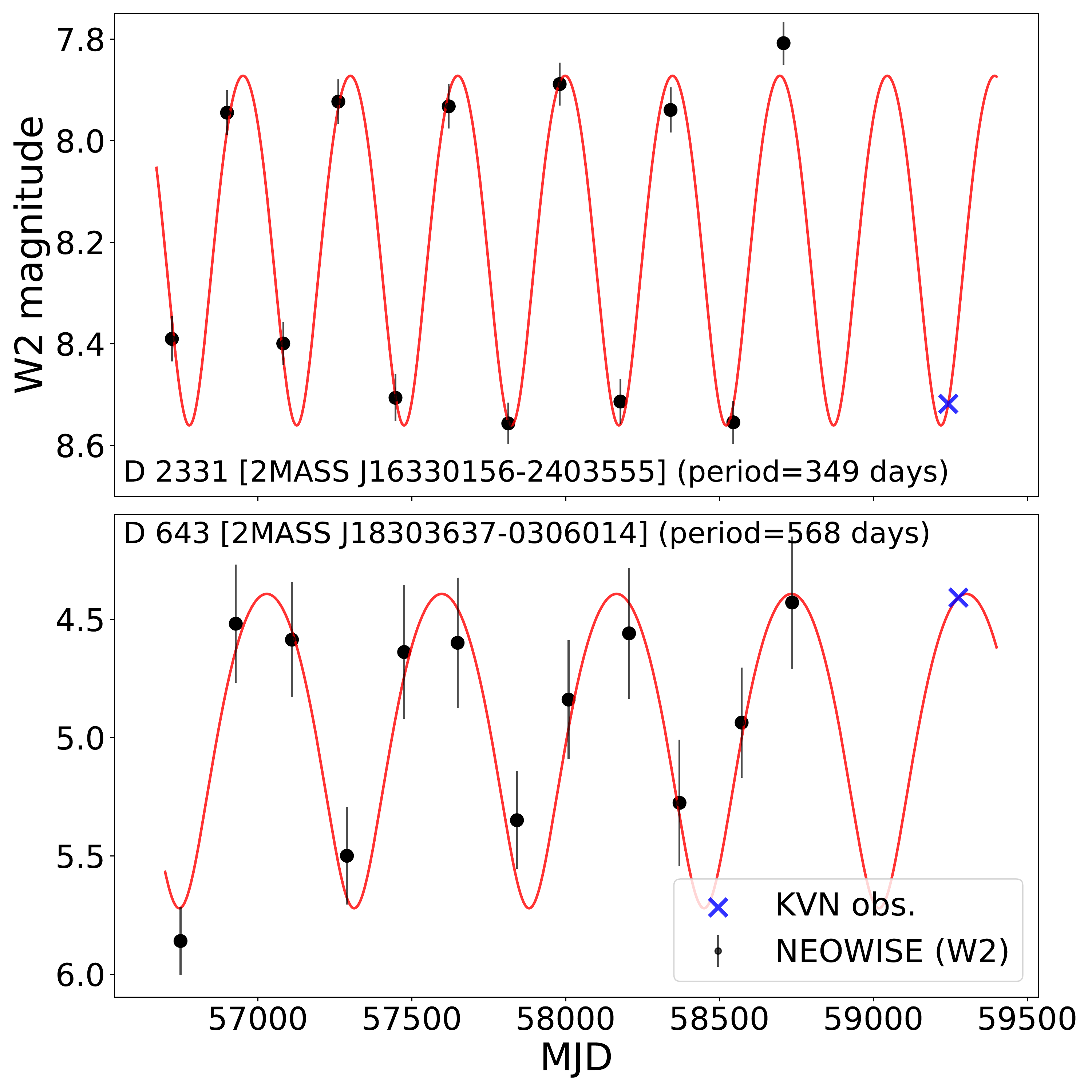}
}
\caption{The NEOWISE light curves of two AGB candidates. The black dots denote the NEOWISE photometric data while the solid red line represents the sinusoidal function found from the periodogram analysis. The source designation and best-fitted period are given at the bottom or top of each panel. The blue crosses indicate the dates when the KVN observations were carried out. 
The lower panel shows the light curves of source D643, where the SiO maser emission was detected. 
\label{fig:neowise}
}
\end{figure}

One potential explanation for these IR sources classified as YSOs (especially Class III) and identified with sinusoidal long period variability would be mis-identification of AGBs to YSOs. Strengthening this argument, in the analysis of Park et al. (submitted), no periodic ``Class III YSOs" have been found in Taurus, where the established YSO catalogue was carefully vetted to ensure no confusion with AGB stars. From the periodogram analysis only with high quality photometry, we identify 56 potential AGB interlopers previously identified as Class III sources. Our target list (Table 1) also includes protostars and disk sources with periodic light curves well fitted by sinusoids. These sources could be actual YSOs, but they could be also potentially AGBs. Indeed, three of these sources are also identified as AGBs in AGB catalogs. 
In total, we consider 77 periodic sources in the periodogram analysis with very well defined sinusoidal NEOWISE light curves as AGB candidates: 3 protostars, 18 disks, and 56 Class III YSOs. These numbers correspond to less than 1\% of protostars and disks with the extracted NEOWISE light curves ($\sim$4200 sources), while it reaches about 5\% of Class III YSOs ($\sim$1200 sources.)

To finalize our target list for the maser observations, we cross-matched the known maser sources (e.g. maserdb.net) with all the YSOs in 8 nearby low mass star forming regions. One source (D31) that is also listed in \citet{Suh17} was detected in the SiO maser emission \citep{Deguchi10} while the source D1184 was detected in the OH maser emission \citep{Eder88}.
Both sources are very bright in the WISE bands to produce large uncertainties ($>$ 0.2 mag) in the photometry, so they were excluded from our initial analyses. 
In order to include such bright sources to our target list, if they have sinusoidal light curves with the periods shorter than 1000 days, we applied the periodogram analysis to the NEOWISE light curves of all catalogue sources with the mean W2 magnitudes brighter than 6 mag without considering their photometric uncertainties. Through this analysis we found 8 additional AGB candidates (2 disks and 6 Class III YSOs). The final number of AGB candidates obtained from the YSO catalogs is therefore 85; however, only 73 sources (Table 1; 5 protostars, 10 disks, and 58 Class III YSOs) were observable with the Korean Very Long Baseline Interferometry Network (KVN) 21-m telescopes during the 2021A observing run. 

Figure 2 shows the color-color diagram of our AGB candidates overlaid on the known AGBs from \citet{Suh17}. The O-rich and C-rich AGB stars can be roughly distinguished on the MIR color-color diagram using photometric data including the ALLWISE data \citep{Suh18}. The green and yellow dots indicate bright C-rich and O-rich AGBs with high-quality WISE data. Most of our targets (red and blue dots) are located in the regime of O-rich AGBs in Figure 2, which includes those targets identified in all four WISE bands. 

\begin{figure*}[t]
\centering
{
\includegraphics[trim={2cm 1cm 4cm 2cm},clip,width=0.98\textwidth]
{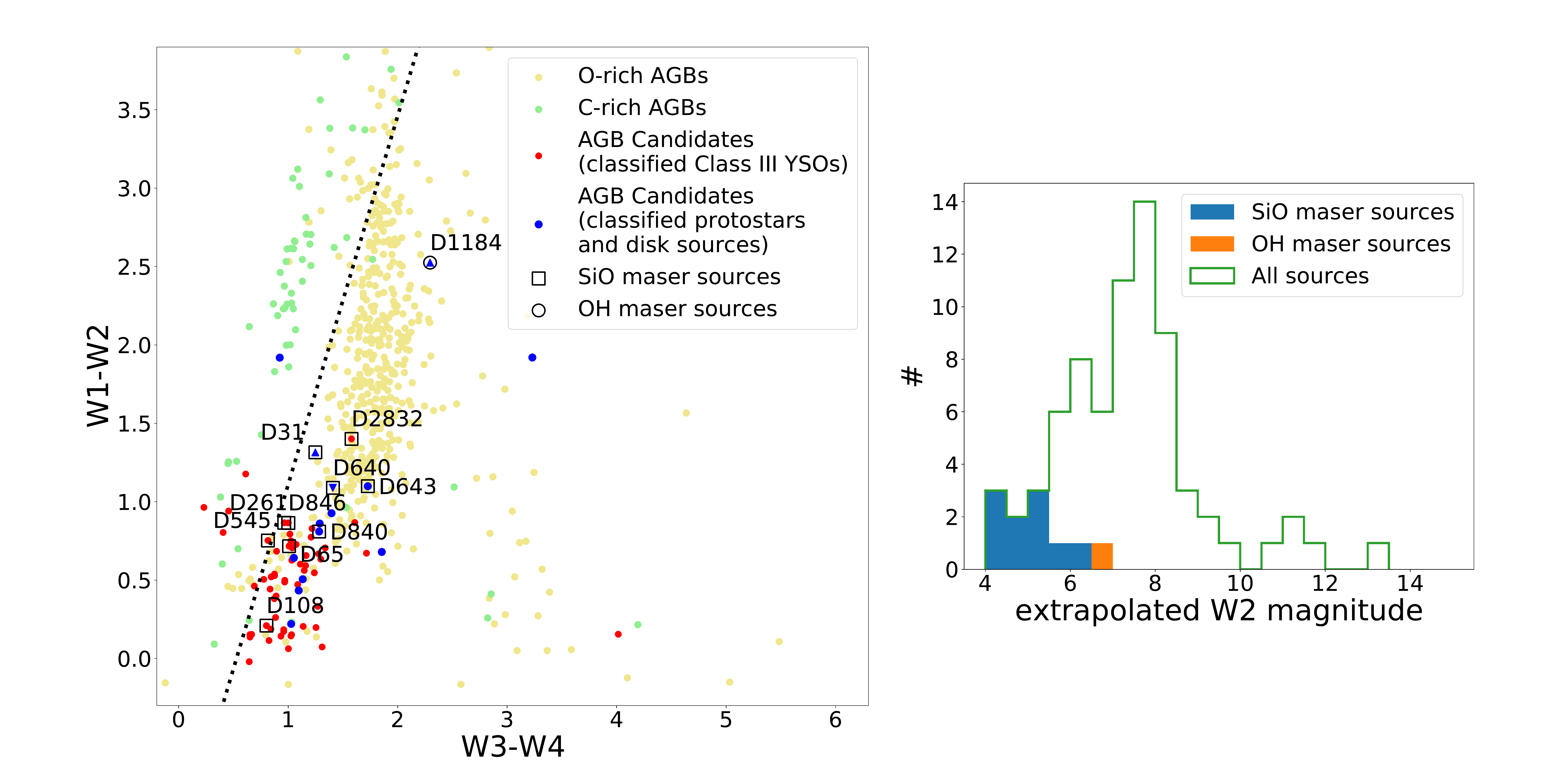}
}
\caption{(Left) The ALLWISE color-color diagram of known AGBs (green and yellow dots) and new AGB candidates (red and blue dots). The C-rich (green dots) and O-rich (yellow dots) AGB stars are roughly divided by the dotted black line \citep{Lian14}. Most of our new AGB candidates are located in the O-rich AGBs regime. The red and blue symbols denote the AGB candidates classified as Class III YSOs and protostars/disk sources, respectively, in the YSO catalogs, and the three blue triangles mark sources listed in both AGB and YSO catalogs. Targets marked with upward triangles are from \citet{Suh17} while the target marked with a downward triangle is from \citet{Lewis20}. 
The open squares indicate AGB candidates detected in the SiO maser emission, while open circles indicate OH maser emission. (Right) The number distribution of the W2 magnitudes extrapolated from the NEOWISE light curves at the observation dates.  All AGB candidates, SiO maser sources, and known OH maser source are presented with the open green, filled blue, and orange histograms, respectively. 
\label{fig:allwise}
}
\end{figure*}   

In the Serpens/Aquila Rift region, the proper motions (PMs) of our AGB candidates are similar to those of background stars rather than the PMs of the YSOs (see Figure 3). 
Here 60 of our AGB candidates were detected by $Gaia$, and 52 sources have the information on their PMs. 
The PMs are divided into two groups: the group for the Serpens star forming region with (PM RA, PM DEC)=(3, $-$8) and the group for the background stars with (PM RA, PM DEC)=($-$3, $-$5), where the unit of PM is mas/yr \citep[see Figure 9 of][]{Herczeg19}. None of our AGB candidates (red and blue dots) are located in the former group, supporting our speculation of AGB contamination in the known YSO catalogues.

 \begin{figure}[t]
\centering
{\includegraphics[trim={0.5cm 0.3cm 0.3cm 0.3cm},clip,width=0.98\columnwidth]
{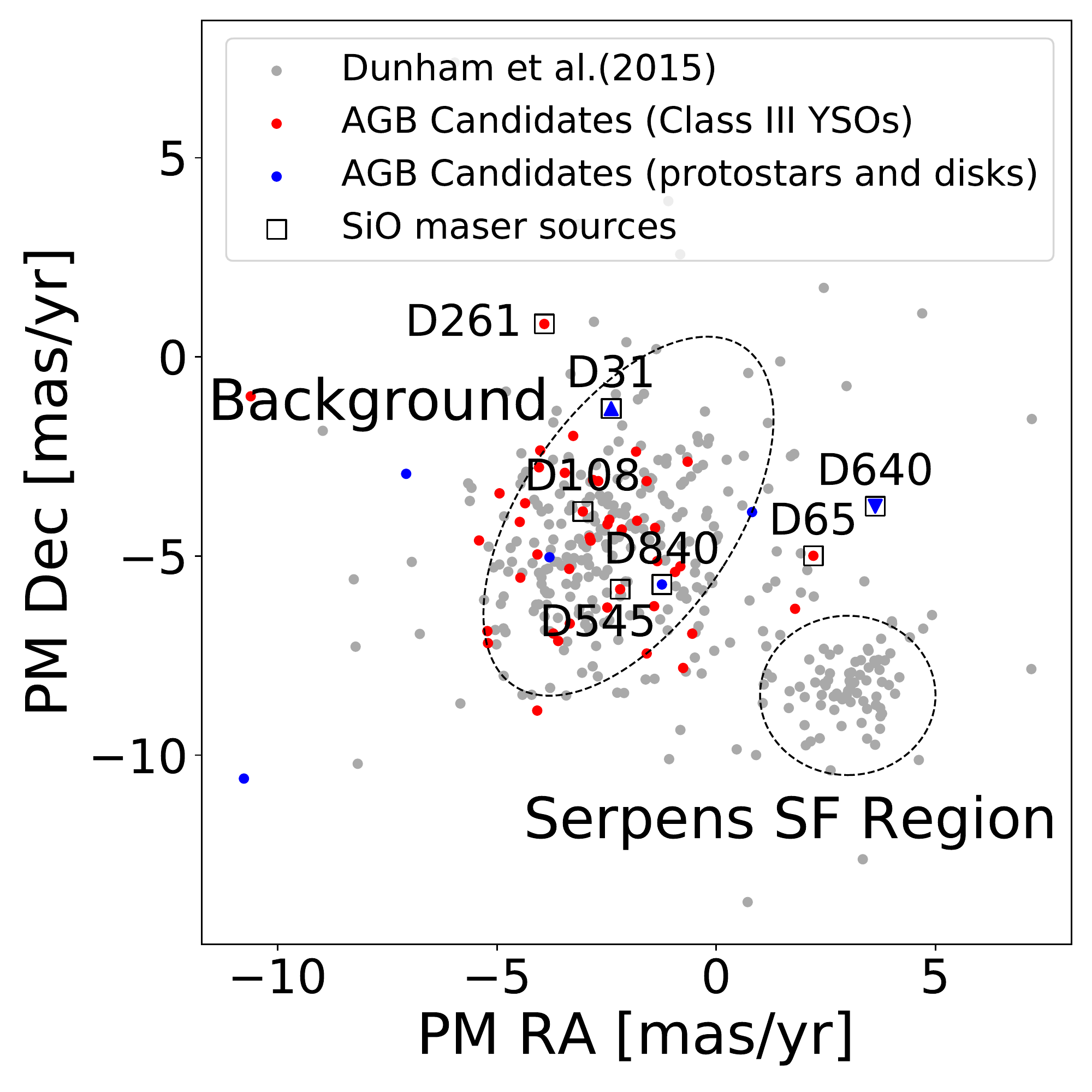}
}
\caption{Proper Motions (PMs) of AGB candidates (red and blue dots) compared to YSOs from the catalog by \citet{Dunham15} (gray dots) in Serpens/Aquila Rift. The astrometric data are from the Gaia Data Release 2 \citep{Gaia18}. We plot only the PMs with their uncertainties smaller than 1 mas yr$^{-1}$ for YSOs while all PMs for the AGB candidates are plotted without considering their uncertainties. PMs are divided into two groups: one group, to the left, consists of background stars while the other group, to the right, consists of YSOs in the Serpens star forming region, as marked with dashed ellipse and circle, respectively  \citep[see Figure 9 in][]{Herczeg19}. Most of our AGB candidates are located in the left group overlapped with background stars. The open symbols, which are the same as in Figure 2, indicate the sources detected in maser emission. Two blue triangles (D31 and D640) indicate the IR sources listed in both AGB and YSO catalogues.
\label{fig:gaia}
}
\end{figure}


\section{KVN observations}

If our targets are indeed O-rich AGBs, the SiO and/or H$_2$O maser emission would be detectable. Variability and SiO maser emission are known to be well correlated, and the SiO masers disappear right after the AGB phase ends \citep{Nyman98}. According to \citet{Lewis89}, both H$_2$O and SiO masers disappear when the mass-loss stops, whereas the OH masers remain active during the post-AGB stage. In this context, we performed the observation of our AGB candidates in the SiO, H$_2$O, and CH$_3$OH maser transitions with the single dish observation mode of the Korean Very Long Baseline Interferometry Network (KVN). The detection of the SiO and/or H$_2$O maser emission can shed light on the nature of the periodic variables with very regular sinusoidal MIR light curves and periods of a few hundreds days since late-stage Class III YSOs never show such maser emission. On the other hand, the methanol maser can distinctly identify actual YSOs \citep{Ladeyschikov19}.

We carried out simultaneous multi-frequency observations of 73 AGB candidates using the 21-meter single-dish telescopes of the KVN  for two months from January 29 to March 30, 2021.
We observed the H$_2$O  6$_{16}$--5$_{23}$ (22.235080 GHz), CH$_3$OH 2$_1$--3$_0$E, 9$_2$--10$_1$A$^+$, 6$_2$--6$_1$E, 7$_0$--6$_1$A$^+$ (19.967396, 23.121024, 25.018112, 44.069476 GHz),  SiO v = 1, 2, 3 J = 1--0 (43.122080, 42.820587, 42.519375 GHz), SiO v = 1, J = 2--1, 3--2 (86.243442 and 129.363359 GHz) maser lines in both right and left circular polarization simultaneously using the multi-frequency receiving system of the KVN \citep{Han08}. A FX-type digital spectrometer is used for 86 and 129 GHz bands and a GPU spectrometer is used for 22 and 43 GHz bands.
The total bandwidths of 16$\times$32 MHz for 22 and 43 GHz bands, and 4$\times$64 MHz for 86 and 129 GHz bands were adopted to provide velocity resolution better than 0.11 km s$^{-1}$ and velocity coverage larger than 140 km s$^{-1}$.
We performed position switching observations, applying a total on-source integration time of about 60 minutes for each object. 
We reach an $rms$ noise level of about 0.01 K for 
all frequencies, with 0.4 km s$^{-1}$ velocity resolution.

Information for our observational set-up is summarized in Table A.1. Typical system temperatures are about 100, 150, 200 and 250 K and average conversion factors from the antenna temperature to the flux density are 12.9, 12.3, 15.9 and 22.8 Jy K$^{-1}$ at 22, 43, 86 and 129 GHz, respectively. 
We conducted sky dipping, pointing and focus observations every one or two hours depending on the sky condition.
Due to a strong spurious feature in the spectra near 20.0 GHz, we could not use the data for CH$_3$OH 2$_1$--3$_0$E  at 19.967396 GHz.

The FWHM beam sizes are 125, 63, 32 and 23 arcseconds in 22, 43, 86 and 129 GHz, respectively.
Since the beam patterns of KVN telescopes have rather high sidelobe level of $\sim$13 dB \citep{Lee11}, a strong maser source located within the sidelobe of a target observation can result in false positive (see Appendix B).
Therefore, we conducted 7$\times$7 grid mapping observations with half beam spacing toward some suspicious detections to confirm that they did not come through the sidelobe. 


\section{Results}

SiO maser emission is detected toward 10 AGB candidates. Table 1 provides a list of sources with detections and the individual spectra are presented in Figure A.1 and Figure  A.2.  H$_2$O and CH$_3$OH maser emission is undetected in our sample.  
The Gaussian fitting results of the detected lines are summarized in Table A.2. 

The upper panel of Figure 4 presents the relation between the maser line strength and the extrapolated MIR brightness at the observed dates, where the extrapolation is taken from the best fit periodic light curve for each source. 
Almost all sources detected in the SiO maser lines are very bright in MIR with the extrapolated W2 brighter than 6 magnitude (see also the right panel of Figure 2).  The one exception, D108, is detected in only one SiO maser line (v=1, J=1--0). In addition, for 8 out of 10 SiO maser sources, the SiO maser emission was detected around the maximum phase of the source light curve (see the bottom panel of Figure 4). 
We cannot entirely rule out the possibility that the W2 magnitude of D108 was brighter than 6.1 mag on the maser observation date since its photometric uncertainty is large.
The sensitivity of our SiO maser survey, therefore, seems sufficient only for targets brighter than 6 mag at W2.

Nine of the 10 SiO maser detections in our sample are first time detections.
As noted earlier, the source D31 was previously detected in the maser lines of SiO J = 1--0,  v = 1 (43.12208 GHz), v = 2 (42.82059 GHz) by \citet{Deguchi10}.
Furthermore, \citet{Eder88} detected the source D1184 in OH N=1$^-$--1$^+$, J = 3/2 -- 3/2, F = 1--2 (1612.23 MHz), although neither H$_2$O nor SiO maser line emission was detected for that source during our observations.
However, we observed D1184 at its extrapolated W2 minimum phase, when the expected W2 magnitude is about 7.5 mag, and thus, apparently too faint to induce detectable SiO maser emission given our sensitivity (see Figure 4). 

As a result, including D1184, in total 11 of our AGB candidates are confirmed as AGBs by the detection of SiO masers and known OH maser emission. This strongly supports our speculation that the MIR sources with the well-fitted sinusoidal light curves and periods of a few hundreds days are very likely AGB interlopers. 

 \begin{figure*}[t]
\centering
{\includegraphics[trim={0cm 2cm 1.5cm 3cm},clip,width=0.7\textwidth]{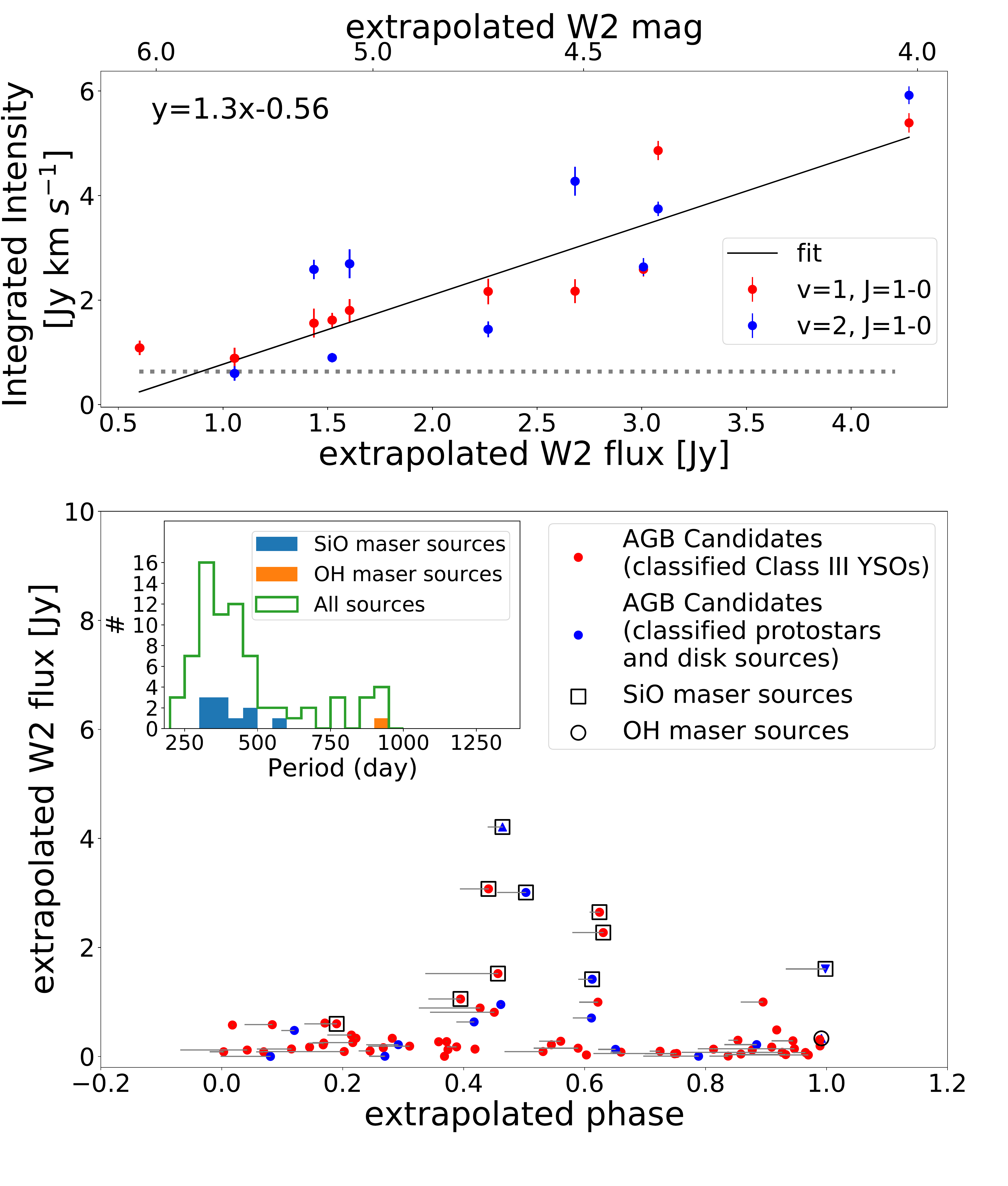}
}
\caption{
The relation between the source MIR brightness and SiO maser strength (top) and the SiO maser detection depending on the extrapolated W2 flux and phase (bottom).
The top panel shows the integrated intensities of the SiO maser lines as functions of the W2 flux, extrapolated to the observing date using the best-fit periodic light curve. The v = 1 and v = 2 lines are fitted together to provide the linear relation (black solid line). The fitted linear relation is presented at the top. The gray dotted horizontal line denote the average sensitivity, 3$\sigma$ (0.633 Jy km s$^{-1}$), of our observations. The bottom panel  presents the extrapolated W2 flux with respective to the source phase at the observing date. The symbols and colors are the same as those in Figures 2 and 3. The horizontal bars indicate the range of phase during the multiple observations. The maximum phase is set as 0.5. The inset presents the histogram of periods determined from the MIR light curves. The colors are the same as those in the inset of Figure 2.
\label{fig:kvn}
}
\end{figure*} 

\section{Discussion}

SiO maser emission is detected toward 10 sources, including two that were previously classified, by \citet{Dunham15}, as Class 0/I, two as Class II, and six as Class III YSOs.
Additionally, the previously known OH maser source was classified as Class 0/I in the YSO catalogs, likely due to its thick circumstellar envelope material. 
Our results clearly demonstrate that the YSO catalogs can be contaminated by AGBs, since SiO masers are very rarely detected in low-mass star forming regions. Only seven SiO maser sources have been detected from YSOs, all  in high-mass star forming regions \citep{Cho16}.

The AGBs confirmed by the SiO detection have periods longer than 300 days, and the period of the OH maser source is $\sim$950 days (see the inset of Figure 4).
The periods and amplitudes obtained from the NEOWISE data for these confirmed AGBs are in the range for typical Mira variables detected at L or M bands \citep[e.g.,][]{Kwon10a, Kwon10b}. All confirmed AGBs are O-rich with relatively bright W2 magnitudes (Figure 2), and 80\% of them were observed around their maximum phases (see the bottom panel of Figure 4). These facts are very consistent with the pumping mechanism of SiO maser emission.

Two main pumping mechanisms for the SiO maser emission have been recognized: radiative pumping \citep{Bujarrabal94} and collisional pumping \citep{Humphreys02}. For the radiative pumping, the infrared radiation at 8 $\mu$m and 4 $\mu$m plays an important role because these wavelengths correspond to the excitation energies of the first and second vibrational states of the SiO molecule, respectively. \citet{Pardo04} showed that the maxima of SiO maser intensities occur near the maximum phases of the near-IR light curves, with a phase lag of 0.05 to 0.20 with respect to optical maxima. This result supports the radiative pumping mechanism of SiO masers rather than the collisional pumping.  The NEOWISE W2 band wavelength (4.6 $\mu$m) is similar to the excitation energy of the second vibrational state, and most of the SiO maser emission was detected around the maximum phases in the W2 brightness of the confirmed AGBs in our survey. 

Long-term monitoring observations of SiO masers for evolved stars present large variations with the stellar pulsation phase \citep[for example, 16.0 to 69.8 Jy in v=1, J=1--0 and 8.5 to 40.6 Jy in v=2, J=1--0 for the minimum to maximum phases of V627 Cas,][]{Yang21}. Therefore, we need to observe all our AGB candidates at the maxima of their W2 light curves, and with a better sensitivity, to confirm their nature more completely.

Half of our SiO maser detected sources show stronger intensity in the v=2, J=1--0 maser line than in the v=1, J=1--0 maser line,
although the v=2 maser requires a higher excitation energy ($\sim$1800 K for v = 1 and $\sim$3600 K for v = 2). 
This result may be associated with the development of a hot dust shell as the AGB stars evolve \citep{Ramstedt12, Cho12}; this tendency has been demonstrated in the survey of post-AGB stars by \citet{Yoon14}. In many post-AGB stars, only the SiO v=2, J=1--0 maser is detected and the v=1 maser is undetected.


The detection rate of SiO masers for our AGB candidates is about 14\% (10 out of 73). Compared to known SiO maser stars,  our detected SiO masers are relatively weak with the peak intensities from 0.21 to 1.77 Jy. \citet{Yoon14} detected about 30\% and 9\% of AGBs and post-AGBs, respectively, in their SiO maser survey with KVN single dish telescopes and interpreted that the detection rate decreases with evolutionary stage.

There were no H$_2$O maser detections among our targets, even in the SiO maser sources. This result may be related to the intensity ratio between SiO and H$_2$O masers and/or different excitation conditions between the two masers, which are associated with the AGB evolution. The intensity of the SiO maser is stronger than that of the H$_2$O maser in most evolved stars. For example, 68\% of 111 Mira variables and 69\% of 32 OH/IR stars showed higher SiO maser peak intensities than those of H$_2$O, with the SiO peak intensity typically more than a factor of 2 stronger than the H$_2$O \citep[see Figure 2 and 3 of][]{Kim14}. 

Alternatively, the outflow property and source mass may play a role in the SiO and H$_2$O maser conditions. \citet{Sevenster02} showed that post-AGBs with high outflow velocities ($\geq 15$ km s$^{-1}$) are located in the left, blue group (``LI" sources) while those with low outflow velocities ($\leq 15$ km s$^{-1}$) are located in the right, red group (``RI" sources) in the IRAS Two-Color Diagram \citep[see Figure 1 of][]{Sevenster02}. For our maser detected sources, 9 out of our 11  have corresponding IRAS sources, and all of them belong to the left, blue group (``LI" sources). 
Previously, SiO masers have been detected toward AGBs located in the LI region, while only H$_2$O masers with no SiO maser detection were detected toward AGBs located in the RI region \citep{Yoon14}. Based on this result, \citet{Yoon14} suggested that more massive O-rich AGB stars in the LI region may enrich heavy elements such as Si, and higher expansion velocities may produce strong shock waves, which in turn form a relatively large number of SiO molecules. 

The distances to our maser-detected sources are not known; however, as they all lie in the direction of Serpens-Aquila, and thus toward the inner Galaxy, they are likely to be associated with the Galactic bulge. This notion is further supported by the observed maser velocities being in the range 40 -- 120\,km\,s$^{-1}$ (see Appendix A), consistent with kinematic distances near the inner Galactic tangent point along that line of sight. Thus, as with the case of the PMs, the source velocities are dissimilar to the $<$10\,km\,s$^{-1}$ LSR velocities of YSOs in the Serpens/Aquila regions. 

Among our sample, the SiO v=1, J=2--1 maser was also detected in 3 sources, each of which also had SiO v=1, J=1--0 maser emission (D65, D545, and D640).
In general, the SiO v=1, J=1--0 maser is expected to be stronger than the SiO v=1, J=2--1 and J=3--2 masers due to its lower rotational energy. 
The peak intensity ratio of J=2--1/J=1--0 is 0.67 and 0.46 in D65 and D545, respectively. However, the ratio is 1.24 in D640. The maser intensity ratios of J=2--1/J=1--0 and J=3--2/J=1--0 are known to have a large spread depending on individual stars and, for a single source, dependent on the stellar pulsation phase. The simultaneous monitoring results of SiO v=1, J=1--0, J=2--1, J=3--2 masers for the Mira variable, TX Cam showed that the average peak intensity ratios among SiO v=1, J=1--0, J=2--1 and J=3--2 masers were 0.85 for J=2--1/J=1--0 and 0.30 for J=3--2/J=1--0, although J=2--1/J=1--0 intensity ratios greater than 1 were also detected in some phases \citep{Cho14}. In addition, the maser line ratios for the O-rich AGB, HaroChavira 20 were 1.33 for J=2--1/J=1--0 and 0.55 for J=3--2/J=1--0 during our observation (see Appendix B).

In summary, 11 out of 73 AGB candidate interlopers among previously classified YSOs were confirmed as AGBs based on our SiO maser detection and a previous OH maser detection. Furthermore, none of our candidates were detected either in the H$_2$O or CH$_3$OH maser. Thus, none of our targets was confirmed as a YSO. We therefore reinforce that care must be taken in the classifications of AGBs and YSOs using infrared color-color or color-magnitude diagrams. Independent discriminators, such as the MIR light curves, proper motion measurements, LSR velocities, and maser observations discussed in this paper, are required for solid confirmation of YSO or AGB status.

\section*{Acknowledgement}

This work was supported by the National Research Foundation of Korea (NRF) grant funded by the Korea government (MSIT) (grant number 2021R1A2C1011718). K.-W.S. was supported by the National Research Foundation of Korea (NRF)
grant funded by the Korea government (MSIT; Ministry of Science and ICT) (No.
NRF-2017R1A2B4002328). S.-H.C. was supported by Basic Science Research Program through the National Research Foundation of Korea (NRF) funded by the Ministry of Education (2019R1I1A1A01059001).
D.J. is supported by the National Research Council of Canada and by an NSERC Discovery Grant. 
G.J.H. is supported by general grant 11773002 awarded by the National Science Foundation of China.

This publication makes use of data products from the Near-Earth Object Wide-field Infrared Survey Explorer (NEOWISE), which is a project of the Jet Propulsion Laboratory/California Institute of Technology. NEOWISE is funded by the National Aeronautics and Space Administration. This research has made use of the NASA/IPAC Infrared Science Archive, which is operated by the Jet Propulsion Laboratory, California Institute of Technology, under contract with the National Aeronautics and Space Administration.

We are grateful to the staff of the KVN who helped to operate the array and to correlate the data. The KVN is a facility operated by the KASI (Korea Astronomy and Space Science Institute). The KVN observations and correlations are supported through the high-speed network connections among the KVN sites provided by the KREONET (Korea Research Environment Open NETwork), which is managed and operated by the KISTI (Korea Institute of Science and Technology Information)

\clearpage

\startlongtable
\begin{deluxetable*}{cccccccccccc}
\tabletypesize{\scriptsize}
\tablecolumns{12}
\tablecaption{AGB candidates\label{tab:table1}}
\tablenum{1}
\tablehead{
\colhead{Source\tablenotemark{a}} & \colhead{RA} & 
\colhead{Dec} & \colhead{Cloud} & 
\colhead{Class} & \colhead{Period} & 
\colhead{Amplitude} & \colhead{Mean W2} & \colhead{Observed Phase}&
\colhead{Extrapolated W2} & \colhead{SiO detection} \\ 
\colhead{} & \colhead{(J2000)} & \colhead{(J2000)} & \colhead{} & 
\colhead{} & \colhead{(days)} & \colhead{(mag)} & 
\colhead{(mag)} & \colhead{} & \colhead{(mag)} & \colhead{}
} 
\startdata
M 669  & 05:38:53.9 & -07:02:33 & Orion     & Class 0/I & 243 & 0.73 & 12.9 & 0.00 - 0.08 & 13.5 &   \\
M 2866 & 05:42:09.3 & -02:09:50 & Orion     & Class II  & 694 & 0.08 & 7.8  & 0.62 - 0.65 & 7.8  &   \\
D 7    & 18:03:13.8 & -04:28:45 & Aquila    & Class III & 948 & 0.10 & 8.2  & 0.06 - 0.07 & 8.3  &   \\
D 31\tablenotemark{b,c} & 18:10:28.9 & -02:37:42 & Aquila & Class 0/I & 485 & 0.53 & 4.5 & 0.44 - 0.46 & 4.0 & Y \\
D 52   & 18:16:31.6 & -03:09:45 & Aquila    & Class III & 406 & 0.07 & 8.9  & 0.70 - 0.75 & 8.9  &   \\
D 57   & 18:21:07.9 & -03:05:33 & Aquila    & Class III & 423 & 0.09 & 9.4  & 0.86 - 0.97 & 9.5  &   \\
D 61   & 18:21:40.3 & -03:18:33 & Aquila    & Class III & 337 & 0.41 & 7.2  & 0.52 - 0.56 & 6.8  &   \\
D 65   & 18:22:11.9 & -03:05:08 & Aquila    & Class III & 423 & 0.32 & 4.7  & 0.39 - 0.44 & 4.4  & Y \\
D 68   & 18:22:21.0 & -02:55:33 & Aquila    & Class III & 406 & 0.35 & 5.3  & 0.86 - 0.90 & 5.6  &   \\
D 82   & 18:23:23.1 & -03:13:50 & Aquila    & Class III & 485 & 0.34 & 7.2  & 0.14 - 0.17 & 7.3  &   \\
D 91   & 18:24:15.0 & -03:00:09 & Aquila    & Class III & 326 & 0.40 & 6.6  & 0.99 - 0.99 & 7.0  &   \\
D 108  & 18:25:59.4 & -03:39:40 & Aquila    & Class III & 375 & 0.35 & 6.1  & 0.14 - 0.19 & 6.1  & Y \\
D 110  & 18:26:04.3 & -03:56:34 & Aquila    & Class III & 796 & 0.10 & 7.4  & 0.91 - 0.91 & 7.5  &   \\
D 118  & 18:26:37.8 & -02:45:51 & Aquila    & Class III & 948 & 0.08 & 8.3  & 0.94 - 0.97 & 8.4  &   \\
D 147  & 18:27:21.1 & -04:11:57 & Aquila    & Class III & 326 & 0.38 & 5.9  & 0.04 - 0.08 & 6.2  &   \\
D 152  & 18:27:23.8 & -03:46:07 & Aquila    & Class III & 375 & 0.04 & 11.8 & 0.37 - 0.37 & 11.8 &   \\
D 182  & 18:27:46.6 & -01:23:30 & Aquila    & Class III & 272 & 0.32 & 7.9  & 0.52 - 0.59 & 7.6  &   \\
D 210  & 18:28:12.0 & -03:34:54 & Aquila    & Class III & 686 & 0.10 & 7.4  & 0.24 - 0.31 & 7.4  &   \\
D 214  & 18:28:20.7 & -04:05:27 & Aquila    & Class III & 442 & 0.39 & 6.2  & 0.34 - 0.45 & 5.8  &   \\
D 261  & 18:28:56.9 & -02:59:55 & Aquila    & Class III & 390 & 0.28 & 4.9  & 0.58 - 0.63 & 4.7  & Y \\
D 274  & 18:29:00.7 & -02:33:30 & Aquila    & Class III & 375 & 0.27 & 7.5  & 0.79 - 0.95 & 7.7  &   \\
D 276  & 18:29:01.1 & -03:34:22 & Aquila    & Class III & 375 & 0.31 & 7.3  & 0.37 - 0.37 & 7.1  &   \\
D 362  & 18:29:18.5 & -01:45:09 & Aquila    & Class III & 265 & 0.38 & 7.5  & 0.88 - 0.88 & 7.7  &   \\
D 391  & 18:29:27.4 & -02:39:48 & Aquila    & Class III & 485 & 0.33 & 6.2  & 0.17 - 0.17 & 6.3  &   \\
D 407  & 18:29:35.8 & -03:38:12 & Aquila    & Class III & 865 & 0.08 & 8.2  & 0.98 - 0.00 & 8.2  &   \\
D 439  & 18:29:42.3 & -03:14:59 & Aquila    & Class II  & 375 & 0.32 & 7.4  & 0.24 - 0.29 & 7.2  &   \\
D 534  & 18:30:06.2 & -02:02:19 & Aquila    & Class II  & 218 & 0.30 & 11.3 & 0.70 - 0.79 & 11.3 &   \\
D 545  & 18:30:08.9 & -01:20:19 & Aquila    & Class III & 316 & 0.26 & 5.4  & 0.34 - 0.46 & 5.1  & Y \\
D 560  & 18:30:13.6 & -02:48:12 & Aquila    & Class III & 423 & 0.34 & 8.5  & 0.47 - 0.53 & 8.2  &   \\
D 632  & 18:30:33.1 & -02:20:58 & Aquila    & Class III & 406 & 0.37 & 7.1  & 0.15 - 0.22 & 7.1  &   \\
D 640\tablenotemark{d}  & 18:30:35.5 & -02:30:35 & Aquila & Class II  & 306 & 0.33 & 4.7 & 0.93 - 1.00 & 5.1 & Y  \\
D 643  & 18:30:36.3 & -03:06:01 & Aquila    & Class 0/I & 568 & 0.66 & 5.1  & 0.46 - 0.50 & 4.4  & Y \\
D 759  & 18:31:16.1 & -01:25:12 & Aquila    & Class III & 297 & 0.32 & 7.8  & 0.27 - 0.27 & 7.7  &   \\
D 780  & 18:31:20.4 & -03:07:40 & Aquila    & Class II  & 375 & 0.29 & 7.0  & 0.83 - 0.88 & 7.2  &   \\
D 786  & 18:31:22.9 & -01:54:24 & Aquila    & Class III & 326 & 0.40 & 6.6  & 0.17 - 0.21 & 6.6  &   \\
D 793  & 18:31:24.9 & -01:52:31 & Aquila    & Class III & 258 & 0.09 & 10.8 & 0.81 - 0.84 & 10.8 &   \\
D 840  & 18:31:36.6 & -01:12:11 & Aquila    & Class II  & 349 & 0.36 & 5.5  & 0.59 - 0.61 & 5.2  & Y \\
D 846  & 18:31:38.5 & -01:50:25 & Aquila    & Class III & 375 & 0.35 & 5.8  & 0.34 - 0.39 & 5.5  & Y \\
D 916  & 18:31:56.2 & -02:41:51 & Aquila    & Class III & 349 & 0.28 & 7.5  & 0.06 - 0.12 & 7.7  &   \\
D 930  & 18:31:59.0 & -02:24:14 & Aquila    & Class III & 362 & 0.58 & 7.4  & 0.81 - 0.81 & 7.5  &   \\
D 932  & 18:32:00.6 & -01:41:15 & Aquila    & Class III & 948 & 0.08 & 6.9  & 0.84 - 0.85 & 6.9  &   \\
D 934  & 18:32:03.5 & -01:10:57 & Aquila    & Class III & 349 & 0.07 & 8.7  & 0.61 - 0.75 & 8.7  &   \\
D 957  & 18:32:13.8 & -01:38:39 & Aquila    & Class III & 375 & 0.33 & 6.0  & 0.33 - 0.43 & 5.7  &   \\
D 994  & 18:32:33.6 & -02:48:18 & Aquila    & Class III & 442 & 0.09 & 8.9  & 0.86 - 0.86 & 8.9  &   \\
D 995  & 18:32:34.0 & -01:58:36 & Aquila    & Class II  & 485 & 0.48 & 6.4  & 0.58 - 0.61 & 6.0  &   \\
D 1001 & 18:32:36.0 & -02:05:24 & Aquila    & Class III & 423 & 0.25 & 7.5  & 0.15 - 0.15 & 7.6  &   \\
D 1008 & 18:32:39.0 & -02:36:12 & Aquila    & Class III & 423 & 0.38 & 6.6  & 0.91 - 0.94 & 6.9  &   \\
D 1012 & 18:32:39.9 & -01:30:09 & Aquila    & Class III & 538 & 0.49 & 7.0  & 0.22 - 0.22 & 7.0  &   \\
D 1026 & 18:32:49.0 & -01:39:19 & Aquila    & Class III & 337 & 0.23 & 6.2  & 0.92 - 0.92 & 6.4  &   \\
D 1049 & 18:33:01.7 & -02:38:54 & Aquila    & Class III & 796 & 0.08 & 8.1  & 0.23 - 0.25 & 8.1  &   \\
D 1061 & 18:33:06.4 & -02:29:05 & Aquila    & Class III & 485 & 0.38 & 5.9  & 0.02 - 0.02 & 6.3  &   \\
D 1064 & 18:33:09.9 & -01:28:59 & Aquila    & Class III & 326 & 0.13 & 7.1  & 0.36 - 0.36 & 7.0  &   \\
D 1074 & 18:33:17.4 & -02:36:25 & Aquila    & Class III & 297 & 0.36 & 7.4  & 0.55 - 0.55 & 7.1  &   \\
D 1078 & 18:33:20.0 & -01:32:05 & Aquila    & Class III & 406 & 0.39 & 7.4  & 0.17 - 0.17 & 7.5  &   \\
D 1080 & 18:33:20.8 & -02:47:57 & Aquila    & Class III & 288 & 0.26 & 8.2  & 0.76 - 0.93 & 8.4  &   \\
D 1127 & 18:36:41.1 & +00:19:14 & Aquila    & Class III & 326 & 0.29 & 7.1  & 0.99 - 0.99 & 7.4  &   \\
D 1128 & 18:36:42.2 & +00:09:15 & Aquila    & Class II  & 569 & 0.16 & 11.5 & 0.24 - 0.27 & 11.4 &   \\
D 1163 & 18:37:33.2 & +00:05:43 & Aquila    & Class III & 316 & 0.32 & 8.1  & 0.42 - 0.42 & 7.8  &   \\
D 1172 & 18:37:42.0 & +00:16:52 & Aquila    & Class III & 272 & 0.10 & 9.5  & 0.60 - 0.60 & 9.4  &   \\
D 1184\tablenotemark{b} & 18:37:55.7 & +00:23:31 & Aquila & Class0/I  & 948 & 0.87 & 5.9 & 0.99 - 0.99 & 6.8 & OH\tablenotemark{e} \\
D 1196 & 18:38:08.6 & +00:20:59 & Aquila    & Class III & 866 & 0.09 & 8.1  & 0.71 - 0.72 & 8.1  &   \\
D 1236 & 18:38:43.5 & +00:01:28 & Aquila    & Class II  & 442 & 0.45 & 6.5  & 0.39 - 0.42 & 6.1  &   \\
D 1242 & 18:38:51.0 & -00:18:52 & Aquila    & Class III & 796 & 0.09 & 7.5  & 0.37 - 0.39 & 7.5  &   \\
D 1247 & 18:38:55.7 & -00:23:40 & Aquila    & Class 0/I & 603 & 0.52 & 6.1  & 0.10 - 0.12 & 6.4  &   \\
D 1279 & 18:39:31.6 & +00:04:57 & Aquila    & Class II  & 510 & 0.36 & 6.0  & 0.46 - 0.46 & 5.6  &   \\
D 1285 & 18:39:35.6 & +00:02:32 & Aquila    & Class III & 349 & 0.13 & 9.1  & 0.83 - 0.93 & 9.2  &   \\
D 1286 & 18:39:40.2 & +00:01:38 & Aquila    & Class III & 316 & 0.30 & 7.7  & 0.93 - 0.04 & 7.9  &   \\
D 1293 & 18:39:54.3 & +00:04:14 & Aquila    & Class III & 375 & 0.32 & 7.0  & 0.28 - 0.28 & 6.9  &   \\
D 1301 & 18:40:11.2 & +00:14:46 & Aquila    & Class III & 485 & 0.35 & 5.9  & 0.59 - 0.62 & 5.6  &   \\
D 1316 & 18:40:25.8 & +00:18:22 & Aquila    & Class III & 240 & 0.22 & 8.0  & 0.37 - 0.37 & 7.9  &   \\
D 2331 & 16:33:01.5 & -24:03:55 & Ophiuchus & Class III & 349 & 0.34 & 8.2  & 0.07 - 0.20 & 8.3  &   \\
D 2809 & 18:29:08.0 & -00:07:37 & Serpens   & Class III & 865 & 0.08 & 8.4  & 0.64 - 0.66 & 8.3  &   \\
D 2832 & 18:29:28.2 & -00:22:57 & Serpens   & Class III & 485 & 0.57 & 5.0  & 0.61 - 0.62 & 4.5  & Y
\enddata
\tablenotetext{a}{Source numbers are the same as those in their original catalogs. M and D are used for the YSOs listed in \citet{Megeath12} and \citet{Dunham15}, respectively.}
\tablenotetext{b}{Targets listed in \citet{Suh17}.}
\tablenotetext{c}{The SiO maser emission was detected by \citet{Deguchi10} as well as our observation.}
\tablenotetext{d}{An AGB star listed in \citet{Lewis20}.}
\tablenotetext{e}{The OH maser emission was detected by \citet{Eder88}}

\end{deluxetable*}

\bibliographystyle{aasjournal}
\bibliography{WISE_AGB}

\clearpage
\appendix

\section{AGB stars confirmed by the SiO masers}
\setcounter{figure}{0}
\renewcommand{\thefigure}{A.\arabic{figure}}

The information on the maser observation is summarized in Table \ref{tab:obs_table}.
The detected SiO maser line spectra are presented in Figure A.1 and Figure A.2, and the Gaussian fitting results of each line are summarized in Table \ref{tab:fit_table}.

\renewcommand\thetable{A.\arabic{table}}
\setcounter{table}{0}

\begin{deluxetable}{ccccccc}[h]
    \tabletypesize{\scriptsize}
    \tablecolumns{7}
    \tablecaption{KVN maser observation\label{tab:obs_table}}
    \tablehead{
    \colhead{Line} & 
    \colhead{Frequency} & \colhead{Bandwidth} & \colhead{Beam Size} & 
    \colhead{Tsys} & 
    \colhead{Velocity Coverage} & \colhead{Conversion Factor}\\ 
    \colhead{} & \colhead{(GHz)} & \colhead{(MHz)} &
    \colhead{(\arcsec)} & \colhead{(K)} & 
    \colhead{(km s$^{-1}$)} & \colhead{(Jy K$^{-1}$)}
    }
    \startdata
    H$_2$O  6$_{16}$--5$_{23}$ & 22.235080 & 512 & 125 & 100 & -170 -- 250 & 12.9 \\
    CH$_3$OH 9$_2$--10$_1$A$^+$& 23.121024 & 512 & 125 & 100 & -170 -- 250 & 12.9 \\
    CH$_3$OH 6$_2$--6$_1$E& 25.018112 & 512 & 125 & 100 & -170 -- 250 & 12.9 \\
    CH$_3$OH 7$_0$--6$_1$A$^+$& 44.069476 & 512 & 63 & 150 & -70 -- 150 & 12.3 \\
    SiO v = 1 J = 1--0& 43.122080 & 512 & 63 & 150 & -70 -- 150 & 12.3 \\
    SiO v = 2 J = 1--0& 42.820587 & 512 & 63 & 150 & -70 -- 150 & 12.3 \\
    SiO v = 3 J = 1--0& 42.519375 & 512 & 63 & 150 & -70 -- 150 & 12.3 \\
    SiO v = 1 J = 2--1& 86.243442 & 256 & 32 & 200 & -110 -- 110 & 15.9 \\
    SiO v = 1 J = 3--2& 129.363359 & 256 & 23 & 250 & -70 -- 70 & 22.8
    \enddata
    \end{deluxetable}

\vspace{0.5cm}

\begin{deluxetable}{ccccccccc}[h]
\tabletypesize{\scriptsize}
\tablecolumns{9}
\tablecaption{Detected SiO masers\label{tab:fit_table}}
\tablehead{
\colhead{Source} &  \colhead{Transition} & 
\colhead{Frequency} & \colhead{rms} & 
\colhead{Velocity} & \colhead{Tpeak} & 
\colhead{FWHM (err)} & \colhead{Integrated Intensity (err)} & \colhead{Observed Phase}\\ 
\colhead{} & \colhead{} & \colhead{(GHz)} & 
\colhead{(Jy)} & \colhead{(km s$^{-1}$)} & \colhead{(Jy)} &
\colhead{(km s$^{-1}$)} & \colhead{(Jy km s$^{-1}$)} & \colhead{}
}
\startdata
D 31   & SiO v=1, J=1-0 & 43.1 & 0.110 & 64.2  & 1.74 & 2.91 (0.11) & 5.39 (0.18) & 0.44 - 0.46   \\
D 31   & SiO v=2, J=1-0 & 42.8 & 0.102 & 64.2  & 2.21 & 2.52 (0.08) & 5.92 (0.17) & 0.44   - 0.46 \\
D 65   & SiO v=1, J=1-0 & 43.1 & 0.093 & 108.5 & 1.42 & 3.22 (0.16) & 4.86 (0.18) & 0.39   - 0.44 \\
D 65   & SiO v=2, J=1-0 & 42.8 & 0.083 & 108.9 & 1.81 & 1.94 (0.11) & 3.74 (0.14) & 0.39   - 0.44 \\
D 65   & SiO v=1, J=2-1 & 86.2 & 0.223 & 108.3 & 0.95 & 2.42 (0.50) & 3.04 (0.40) & 0.39   - 0.44 \\
D 108  & SiO v=1, J=1-0 & 43.1 & 0.103 & 99.9  & 0.47 & 2.16 (0.29) & 1.09 (0.14) & 0.14   - 0.19 \\
D 261  & SiO v=1, J=1-0 & 43.1 & 0.120 & 60.5  & 0.46 & 4.39 (0.58) & 2.16 (0.25) & 0.58   - 0.63 \\
D 261  & SiO v=2, J=1-0 & 42.8 & 0.123 & 62.3  & 0.85 & 1.59 (0.22) & 1.44 (0.15) & 0.58   - 0.63 \\
D 545  & SiO v=1, J=1-0 & 43.1 & 0.101 & 78.3  & 1.12 & 1.36 (0.02) & 1.62 (0.14) & 0.34   - 0.46 \\
D 545  & SiO v=2, J=1-0 & 42.8 & 0.087 & 78.4  & 0.68 & 1.24 (0.15) & 0.90 (0.09) & 0.34   - 0.46 \\
D 545  & SiO v=1, J=2-1 & 86.2 & 0.173 & 77.5  & 0.52 & 3.56 (0.75) & 2.04 (0.34) & 0.34   - 0.46 \\
D 640  & SiO v=1, J=1-0 & 43.1 & 0.136 & 77.4  & 0.61 & 2.76 (0.38) & 1.80 (0.22) & 0.93   - 1.00 \\
D 640  & SiO v=2, J=1-0 & 42.8 & 0.133 & 77.3  & 0.62 & 4.10 (0.48) & 2.70 (0.28) & 0.93   - 1.00 \\
D 640  & SiO v=1, J=2-1 & 86.2 & 0.239 & 76.7  & 0.76 & 2.19 (0.55) & 1.74 (0.36) & 0.93   - 1.00 \\
D 643  & SiO v=1, J=1-0 & 43.1 & 0.075 & 58.2  & 0.78 & 3.10 (0.19) & 2.59 (0.14) & 0.46   - 0.50 \\
D 643  & SiO v=2, J=1-0 & 42.8 & 0.074 & 58.3  & 0.49 & 5.09 (0.38) & 2.63 (0.17) & 0.46   - 0.50 \\
D 840  & SiO v=1, J=1-0 & 43.1 & 0.119 & 43.7  & 0.26 & 5.63 (1.20) & 1.56 (0.28) & 0.59   - 0.61 \\
D 840  & SiO v=2, J=1-0 & 42.8 & 0.123 & 43.3  & 1.12 & 2.17 (0.17) & 2.59 (0.18) & 0.59   - 0.61 \\
D 846  & SiO v=1, J=1-0 & 43.1 & 0.131 & 37.6  & 0.47 & 1.77 (1.74) & 0.89 (0.20) & 0.34   - 0.39 \\
D 846  & SiO v=2, J=1-0 & 42.8 & 0.121 & 37.7  & 0.45 & 1.24 (0.38) & 0.60 (0.14) & 0.34   - 0.39 \\
D 2832 & SiO v=1, J=1-0 & 43.1 & 0.150 & 116.1 & 0.89 & 2.25 (0.30) & 2.17 (0.23) & 0.61   - 0.62 \\
D 2832 & SiO v=2, J=1-0 & 42.8 & 0.149 & 116.8 & 1.02 & 3.92 (0.28) & 4.27 (0.28) & 0.61   - 0.62
\enddata
\end{deluxetable}

\pagebreak

\begin{figure*}[h!]
\centering
{
\includegraphics[trim={3cm 2cm 2cm 0cm},clip,width=1\columnwidth]
{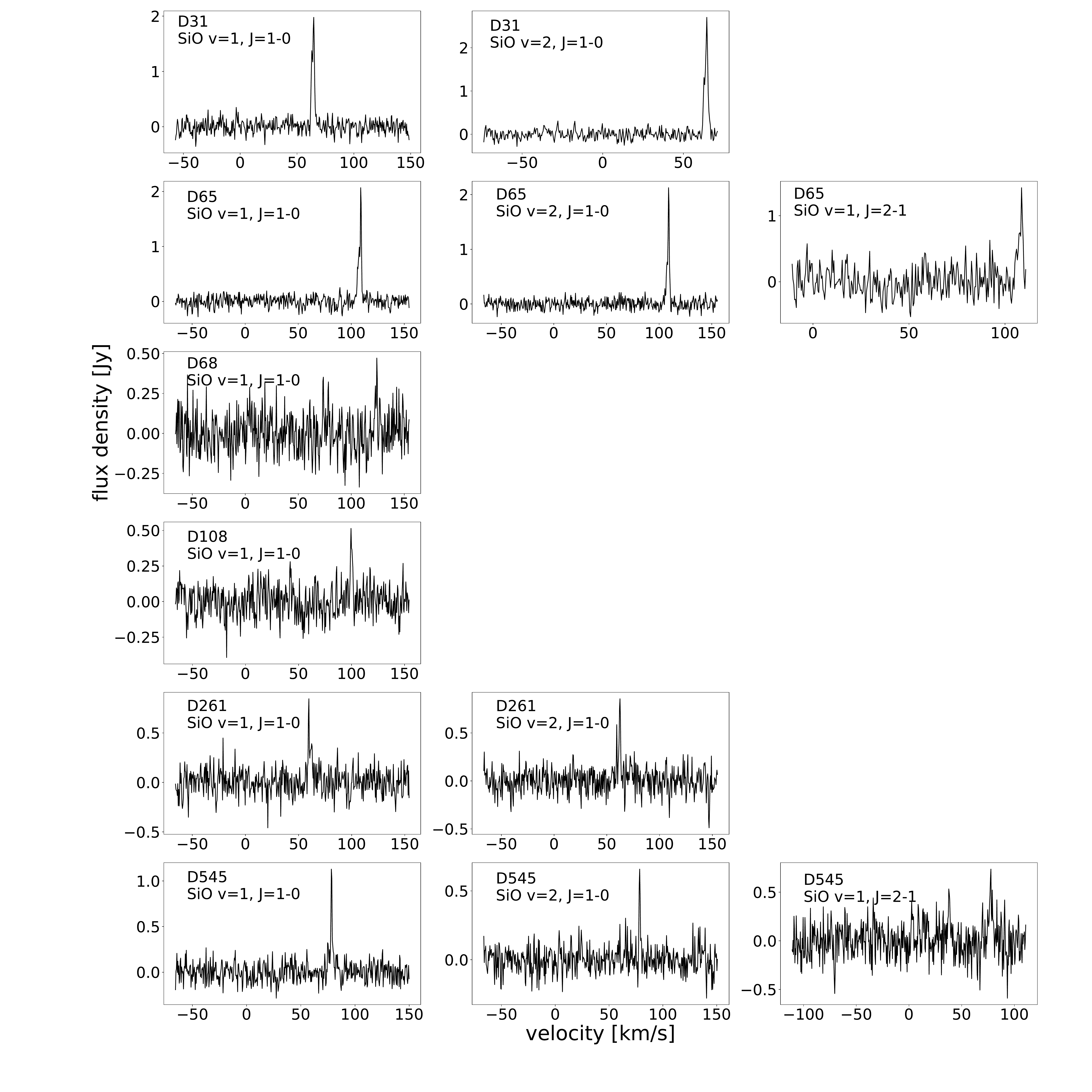}
}

\caption{The maser lines detected in the AGB candidates.  
\label{fig:spec1}
}
\end{figure*}

\begin{figure*}[h!]
\centering
{\includegraphics[trim={3cm 1cm 2cm 0cm},clip,width=1\columnwidth]
{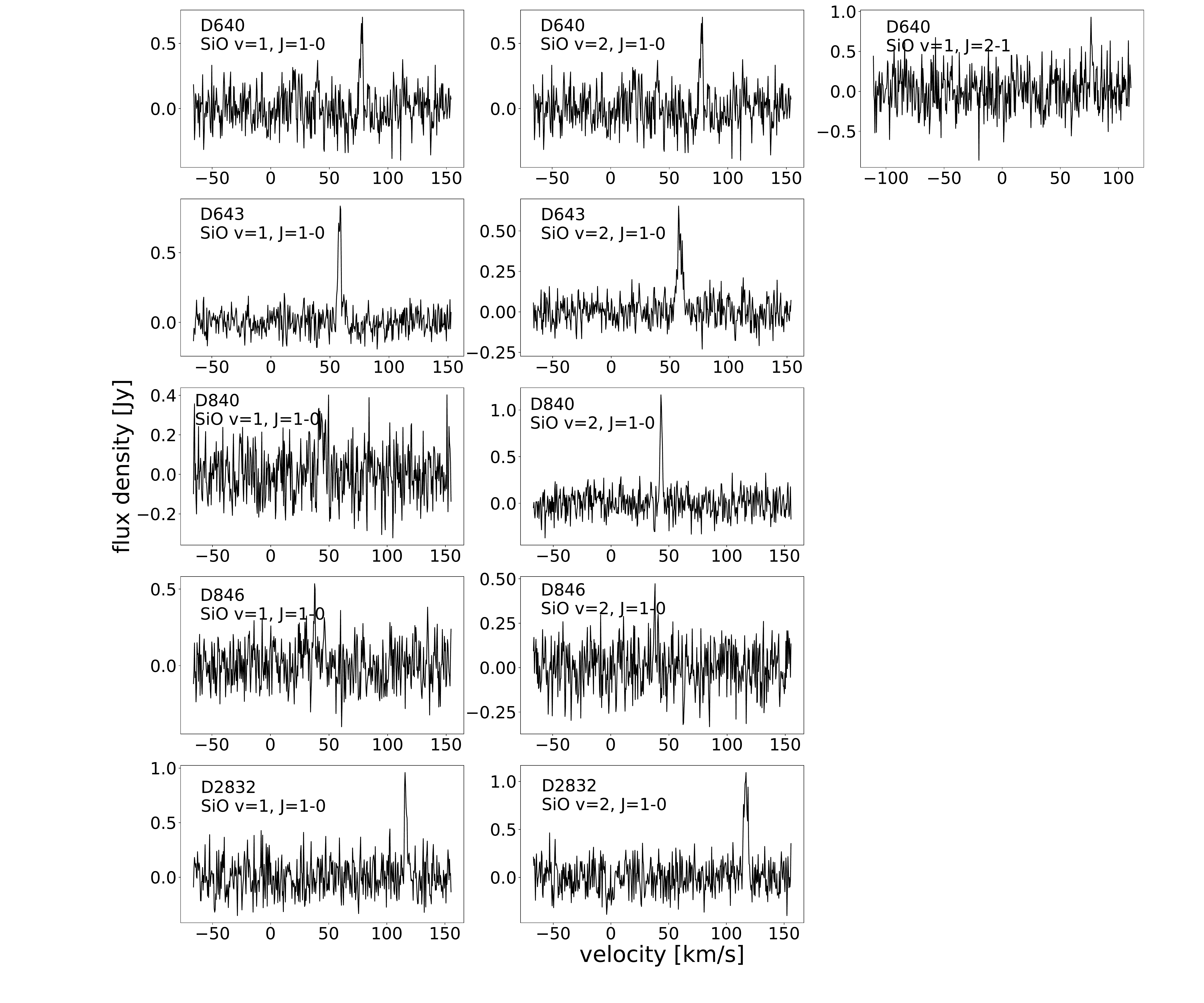}
}
\caption{The maser lines detected in the AGB candidates.
\label{fig:spec2}
}
\end{figure*}


\pagebreak

\section{HaroChavira 20}
\setcounter{figure}{0}
\renewcommand{\thefigure}{B.\arabic{figure}}

\renewcommand\thetable{B.\arabic{table}}
\setcounter{table}{0}

\begin{figure}[h]
\centering
{\includegraphics[trim={5cm 4.5cm 2cm 3cm},clip,width=1\columnwidth]
{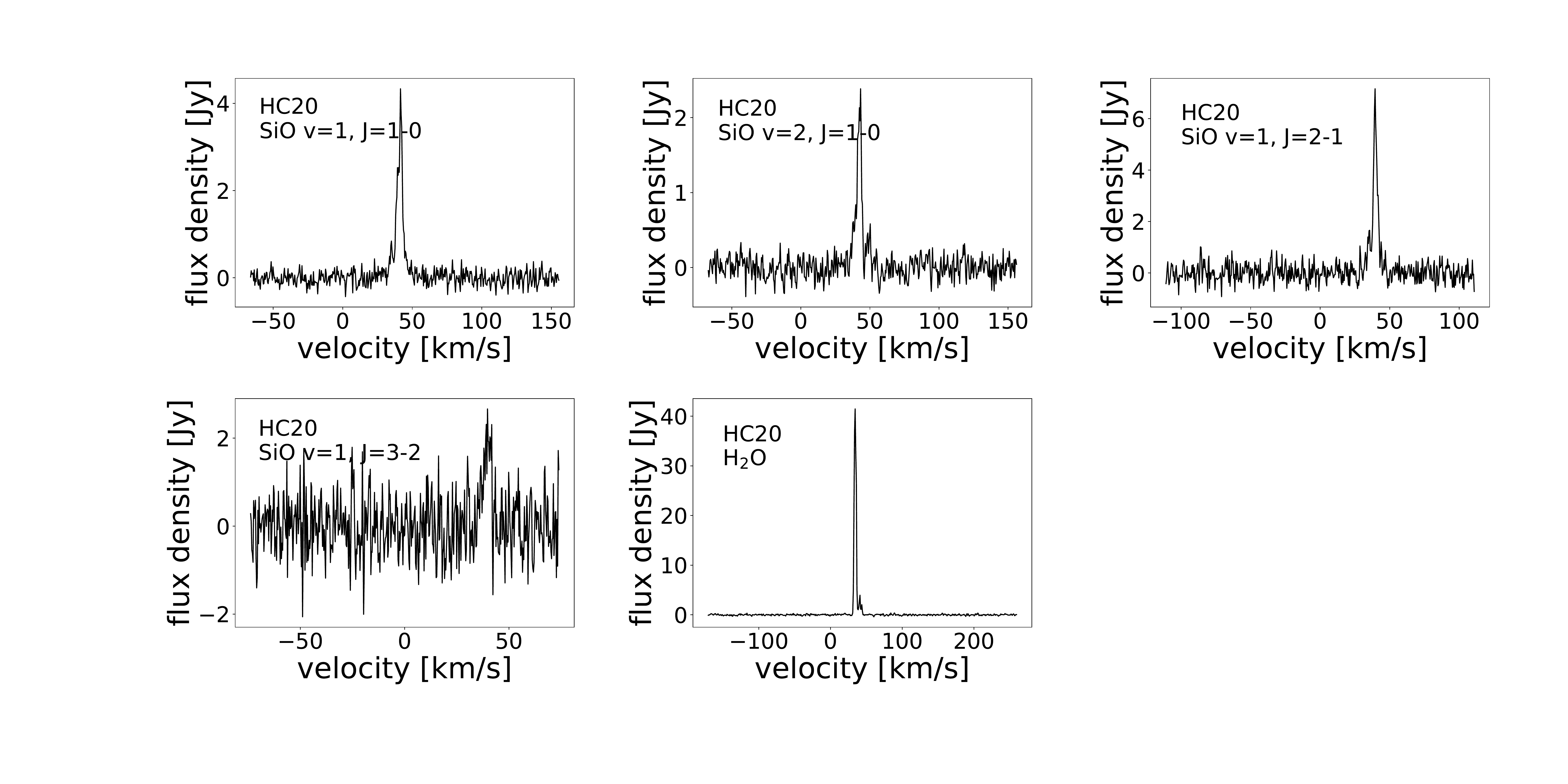}
}
\caption{The maser lines detected in HaroChavira 20.
\label{fig:hc20_kvn}
}
\end{figure}

\begin{figure}[h]
\centering
{\includegraphics[trim={0.5cm 0.3cm 0.3cm 0.3cm},clip,width=0.6\columnwidth]
{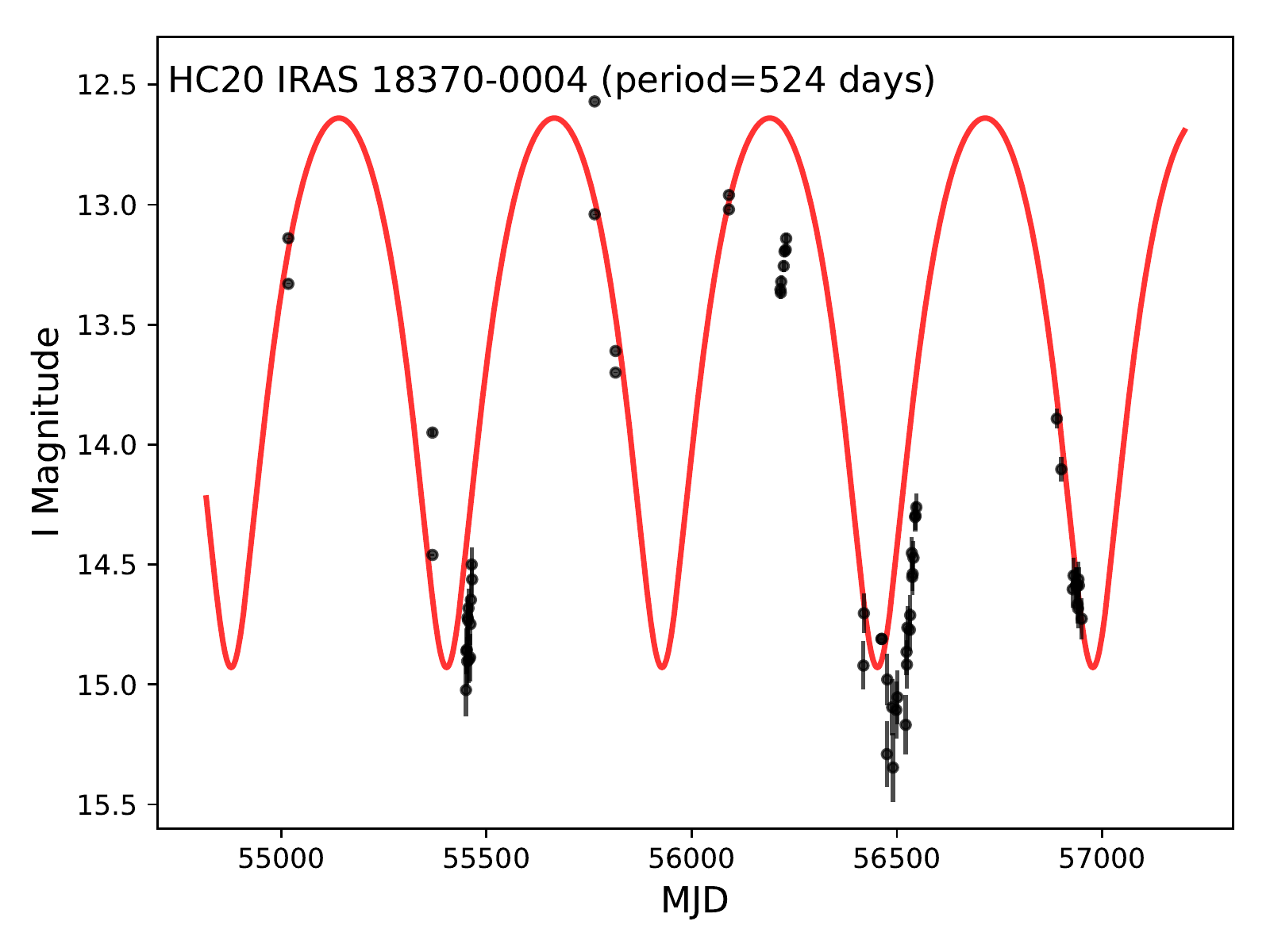}
}
\caption{The I band light curve of HaroChavira 20. The best-fitted sinusoid is presented by the red line. The observation of the maser lines was carried out at the maximum phase if the fitted period is correct.
\label{fig:hc20_vvv}
}
\end{figure}

We detected H$_2$O maser emission toward sources D1285 and D1286 with our single pointing observations. These two sources are only 1.5\arcmin\ apart and suspiciously the maser detections had the same intensity and central velocity. 
A strong maser source can interfere with measurements of nearby sources because the sidelobes of the KVN beam are large. In order to check this possibility, we mapped the $7\arcmin \times7 \arcmin$ area around the source D1286 and found a very strong water maser source located $\sim 4\arcmin$ to the south. The source is HaroChavira 20 (also known as IRAS 18370-0004 or G031.5679+02.6049; d=1.66 kpc) and was previously detected in the SiO v=1 and 2 (J=1--0) maser lines in 2003 \citep{Deguchi04}. 
Toward HarcoChavira 20 we also detected the SiO J=2--1 and J=3--2 (v=1) lines, as well as the 43 GHz SiO maser lines, all of which are presented in Figure \ref{fig:hc20_kvn} and Table \ref{tab:hc20_table}.
The water maser line in this source was detected for the first time in our observation.
HaroChavira 20 has been classified as an O-type AGB star, which has magnitudes of 13.34, 6.01, 4.01, 3.07, and 0 in the G, J, H, K, and W2 bands.
HaroChavira 20 has been covered by the VISTA Variables in the Via Lactea (VVV) survey \citep{Minniti10}, and Figure \ref{fig:hc20_vvv} shows the I-band light curve (black dots) and the best-fitted sinusoid (red line) with the period of 524 days.


\begin{deluxetable}{ccccccc}[h]
\tabletypesize{\scriptsize}
\tablecolumns{7}
\tablecaption{Detected maser lines in HaroChavira 20\label{tab:hc20_table}}
\tablehead{
\colhead{Transition} & 
\colhead{Frequency} & \colhead{rms} & 
\colhead{Velocity} & \colhead{Tpeak} & 
\colhead{FWHM (err)} & \colhead{Integrated Intensity (err)}\\ 
\colhead{} & \colhead{(GHz)} & 
\colhead{(Jy)} & \colhead{(km s$^{-1}$)} & \colhead{(Jy)} &
\colhead{(km s$^{-1}$)} & \colhead{(Jy km s$^{-1}$)}
}
\startdata
SiO v=1, J=1-0    & 43.1                & 0.154     & 41.2                                      & 3.39           & 4.86 (0.13)                                & 17.54 (0.37)                                                  \\
SiO v=2, J=1-0    & 42.8                & 0.138     & 42.3                                      & 2.09           & 4.01 (0.19)                                & 8.93 (0.32)                                                  \\
SiO v=1, J=2-1    & 86.2                & 0.356     & 39.7                                      & 6.14           & 3.56 (0.15)                                & 23.28 (0.74)                                                  \\
SiO v=1, J=3-2    & 129.3                & 0.633     & 39.5                                      & 2.03           & 4.48 (0.61)                                &  9.67 (1.08)                                                  \\
H$_2$O 6$_{16}$–5$_{23}$   & 22.2                & 0.284     & 34.4                                      & 42.54           & 3.28 (0.01)                                & 148.65 (0.66)      
\enddata
\end{deluxetable}

\end{document}